# Conversational Swarms of Humans and AI Agents enable Hybrid Collaborative Decision-making

Louis Rosenberg
Unanimous AI
Pismo Beach, California
Louis@Unanimousa.ai

Hans Schumann
Unanimous AI
San Francisco, California
Hans@Unanimous.ai

Christopher Dishop
Carnegie Mellon University
Pittsburgh, Pennsylvania
cdishop@andrew.cmu.edu

Gregg Willcox
Unanimous AI
Seattle, Washington
Gregg@Unanimous.ai

Anita Woolley
Carnegie Mellon University
Pittsburgh, Pennsylvania
awoolley@andrew.cmu.edu

Ganesh Mani
Carnegie Mellon University
Pittsburgh, Pennsylvania
ganeshm@andrew.cmu.edu

*Abstract*— Conversational Swarm Intelligence (CSI) is an AI-powered communication and collaboration technology that allows large, networked groups (of potentially unlimited size) to hold thoughtful conversational deliberations in real-time. Inspired by the efficient decision-making dynamics of fish schools, CSI divides a human population into a set of small subgroups connected by AI agents. This enables the full group to hold a unified conversation. In this study, groups of 25 participants were tasked with selecting a roster of players in a real Fantasy Baseball contest. A total of 10 trials were run using CSI. In half the trials, each subgroup was augmented with a fact-providing AI agent referred to herein as an Infobot. The Infobot was loaded with a wide range of MLB statistics. The human participants could query the Infobot the same way they would query other persons in their subgroup. Results show that when using CSI, the 25-person groups outperformed 72% of individually surveyed participants and showed significant intelligence amplification versus the mean score (p=0.016). The CSI-enabled groups also significantly outperformed the most popular picks across the collected surveys for each daily contest (p<0.001). The CSI sessions that used Infobots scored slightly higher than those that did not, but it was not statistically significant in this study. That said, 85% of participants agreed with the statement "Our decisions were stronger because of information provided by the Infobot," and only 4% disagreed. In addition, deliberations that used Infobots showed significantly less variance (p=0.039) in conversational content across members. This suggests that Infobots promoted more balanced discussions in which fewer members dominated the dialog. This may be because the infobot enabled participants to confidently express opinions with the support of factual data.

## I. INTRODUCTION

It is well established in the field of Collective Intelligence (CI) that human groups can make collaborative estimations, decisions and forecasts that with accuracy that beats individual participants [1, 16, 17]. Common techniques involve collecting and aggregating data from individual members. This is often described as harnessing the "Wisdom of Crowds" (WoC) Such methods can enhance decision-making but are often limited to narrow tasks such as numerical estimations and fixed-choice selections [2]. In many real-world domains, problems are far more complex, involving a wide range of competing factors in which difficult tradeoffs must be weighed. As a consequence, most traditional CI methods are often not easily applied to group decision-making on complex, real-world, open-ended problems.

Conversational Swarm Intelligence (CSI) is a relatively new CI methodology and technology that aims to address the limitations of traditional methods by enabling large, networked groups (of potentially unlimited size) to hold thoughtful real-time conversational deliberations and converge on solutions that increase collective intelligence [3-6]. In the sections below, the basic principles of CSI are reviewed, and a new feature called an "infobot" is introduced. In addition, an academic study is described in which CSI technology, with and without the infobot feature, is tested in a real-world forecasting task – the selection of a roster of players in a daily Fantasy Baseball contest. This task was chosen because it offers a complex, open-ended scenario in which participants must make tradeoffs to balance competing factors within a fixed budget. Specifically, groups are required to strategically allocate their budget and manage tradeoffs by collectively deliberating about how much to spend on particularly players on their team. We compare the performance of groups reaching collective decisions via traditional survey aggregation against groups deliberating using an online CSI platform called Thinkscape™, with and without the use of Infobots. We also collect subjective feedback about the participants' experience deliberating using CSI technology and about the perceived value of Infobots.

## II. CONVERSATIONAL SWARM INTELLIGENCE (CSI)

CSI is a new collaboration, communication, and collective intelligence technology that enables large, networked groups of individuals to engage in real-time conversations online. As implemented in current tools like Thinkscape, hundreds of individuals can hold thoughtful deliberations in which they collectively brainstorm ideas, debate alternatives, prioritize options, estimate outcomes and converge on unified solutions that have been shown to amplify collective intelligence [6-9]. Current CSI platforms support text-based conversations (with optional voice-to-text). Future platforms could use the same method to enable videoconferencing at a very large scale.

Enabling productive and thoughtful conversations at scale is a longstanding problem that has lacked an effective solution. Although large groups can congregate in online chatrooms or videoconferencing platforms, deliberating at scale is rarely effective, as conversational quality degrades with group size. Research shows that when groups grow beyond 4 to 7 individuals, deliberations tend to become less effective. Each person gets less time to speak, they have to wait longer to respond to others, and often a few dominant personalities control the conversation and sway the full group in a biased direction [10-12].

To solve this, CSI technology learns from Mother Nature by employing Swarm Intelligence techniques modeled on the rapid and efficient decision-making of fish schools [13]. Large schools can make effective life-or-death decisions extremely quickly despite the fact that (i) each individual has a very limited view of the world around them and the real-time threats they face (i.e., partial information) and (b) the group has no central authority that oversees the decision-making process. Evolution developed an efficient solution by enabling each member of the school to "deliberate" in real-time with a small set of nearby fish (i.e. neighbors) using a specialized organ called a *lateral line*. This unique organ detects pressure changes and vibrations in the water caused by the movements of their neighbors. Since each small groups overlap with other small groups, the local decisions propagate across subgroups, enabling information to quickly spread through the full population. This allows thousands of fish to rapidly converge on unified decisions in real-time [14].

The technology of CSI emulates the distributed deliberative process of large fish schools by splitting large human groups into a network of overlapping subgroups, each one containing 4 to 7 participants for optimal real-time conversational deliberation. This alone does not solve the problem, however. That's because humans do not possess the fish-like ability to deliberate within overlapping subgroups. In fact, we find it extremely challenging to participate effectively in more than one conversation at a time. Researchers often refer to this human limitation as the "cocktail party problem" because parties are a very common venue where small conversational subgroups congregate in close proximity. In such settings, if a person's attention shifts to a neighboring conversation instead of the one they are in, their ability to follow both conversation quickly diminishes. In fact, humans evolved the oppositive capability – we are quite skilled at tuning-out neighboring conversations so we can focus intently on local conversation we are engaged in. [14]

To enable overlapping conversational groups of human users in a manner that does not confuse or distract the participants, Conversational Swarm Intelligence employs a novel artificial agent referred to as "Surrogate Agent" that is enabled by real-time API access to Large Language Models (LLMs) [3-9, 14, 15]. CSI achieves this by automatically dividing a large group into a network of small subgroups and inserting a Surrogate Agent into each subgroup. The Surrogate Agent is tasked with observing the local deliberation of the group it is in, distilling the real-time content, and efficiently passing critical points to other subgroups where its local Surrogate Agent will express those points as natural dialog within that local conversation. This enables conversational content to efficiently propagate across subgroups. In this way, CSI empowers groups of potentially any size to hold thoughtful real-time conversational deliberations in which they can share opinions, debate options, brainstorm alternatives, discuss challenges, prioritize competing factors, and converge on solutions that maximize groupwise support.

Figure 1 shows a CSI structure in which a large group of 100 participants divided into 14 subgroups. As shown, each subgroup is populated with an AI-powered Surrogate Agent. While the diagram only shows network connections between pairs of subgroups, the Thinkscape CSI platform used in this study employed a *fully connected* network architecture among subgroups. Using this structure, the Surrogate Agent within each subgroup can pass or receive content from any other subgroup in the network, mediated by *a matchmaking subsystem* that intelligently considers the real-time conversational content in all subgroups in real-time and passes content between groups in a manner that maximize the mixing of diverse insights. In addition, the CSI structure is inherently scalable. This means it could be used to connect hundreds, thousands, or potentially even millions of individual participants in real-time, either using a flat architecture (as shown below) or a nested structure in which a hierarchy is used in the network architecture that creates subgroups of subgroups.

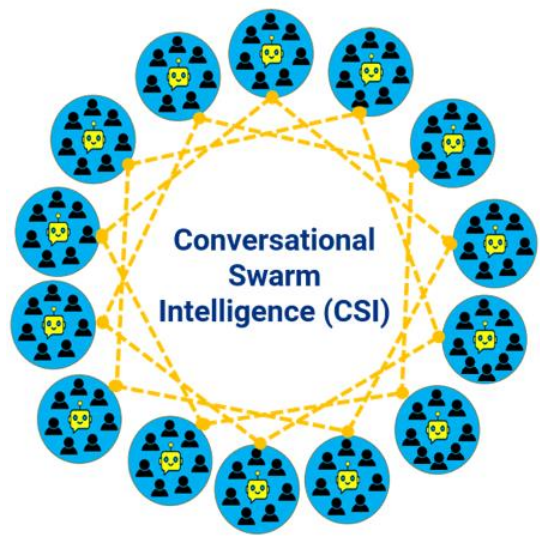

Fig. 1. CSI structure with 100 participants in 14 subgroups with AI agents assigned to each subgroup for passing and receiving conversational content.

Recent studies have shown that CSI enables participants to harness their combined knowledge, wisdom, and insights to solve problems in real-time. In addition, CSI reduces the biasing influence of strong personalities (i.e. loudmouths) because each individual is only directly exposed to a small subgroup of others in real-time. For an insight to propagate beyond its originating subgroup, it must be compelling on its own merits. Similarly, CSI combats the biasing influence of early comments swaying the full group, as if often the problem in traditional focus groups. In the diagram above, different "early comments" would emerge in all 14 subgroups and would spread across the network based on their merits, not the timing of when they were introduced.

In addition, the CSI structure promotes greater dialog per person than traditional methods. In one recent study conducted at Carnegie Mellon, groups of approximately 50 participants were tasked with debating a "current events" topic. When connected using CSI, the individual participants expressed 51% more conversational content on average (p<0.001) compared to trials using traditional centralized chat. In addition, the groups using CSI displayed 37% less variance in contribution between the most vocal and least vocal participants. This suggests that CSI promotes more balanced deliberations. In addition, the majority of the human subjects tested reported preferring the CSI platform over standard centralized chat (p<0.05) and reported feeling they had greater impact on the conversation when they used the CSI platform (p<0.01) [6].

In another recent study, 245 participants were challenged with estimating the quantity of gumballs in a large glass jar. When deliberating together in real-time using Thinkscape CSI platform, the group of 245 participants were automatically split into 47 subgroups of 5 or 6 human members and one AI-powered Surrogate Agent. The groupwise estimations generated using the Thinkscape CSI platform produced a 50% smaller error than the traditional survey-based WoC method [9].

In another recent experiment, groups of 35 people were challenged with answering IQ test questions from standard battery. The study compared individual survey performance on the IQ (as individuals and by WoC aggregation) versus groupwise decisions emerging from conversational deliberations in a CSI-based conversational platform. In all cases, participants were allotted the same time per question. Compared to the average individual IQ score of 100 (the 50th percentile), CSI groups scored an IQ of 128 (97th percentile) which is considered "gifted" by most criteria. This significantly outperformed the WoC aggregation which scored IQ of 115 (84th percentile). In addition, not a single participant that was tested in this study scored a personal IQ that was as high as the groups deliberating together using the CSI platform [15]. This result suggests that CSI may be a possible pathway to enabling very large groups to form a collective superintelligence.

The forecasting study described below tested a new intervention – an additional conversational AI agent referred to herein as an "Infobot" was deployed and tested within groupwise deliberations for the first time. As implemented, an Infobot is primed with specific factual information regarding the task or problem at hand and added to each subgroup (independently). Each Infobot is designed to respond to factual queries from members of its subgroup, providing factual information that is strictly limited to the topic at hand and the set of factual data it was primed with. The infobots are conversational. They are designed to participate in each subgroup alongside human participants. The only difference is that infobots can only respond to direct queries and only give factual responses.

Figure 2 below shows a diagram of a CSI system with Infobots primed with factual and statistical information about MLB players and teams. As shown, this novel structure enables <u>hybrid deliberation</u> in which groups of human users can collaboratively discuss a problem while having an informational AI agents seamlessly bring factual content into the deliberation for human participants to consider and respond to. In addition, because the CSI structure is distributed discussion, it efficiently uses the Infobot concept such that every local group can explore different factual information in parallel, the impact of which can propagate across the full network using standard CSI methods.

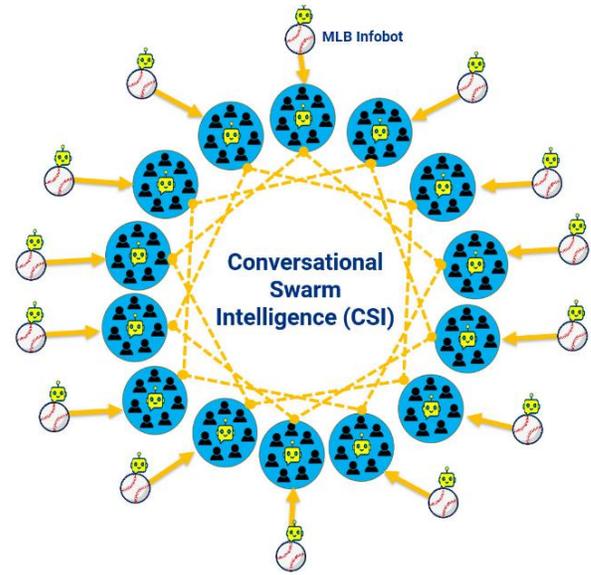

Fig 2. Conversational Swarm Intelligence (CSI) with Infobots.

### III. MLB Fantasy – Group Deliberation Study

To verify the effectiveness of CSI technology in groupwise deliberations of complex problems with competing factors that demand tradeoffs (with and without the use of Infobots), the Thinkscape CSI platform developed by Unanimous AI was used to conduct a collaborative challenge based on a common Fantasy Baseball contest. To capture baseline data, participants were also tasked with creating their own personal roster using a standard survey. The task required participants to "purchase" a set of **5** players to (bi-weekly) for their team while staying within a fixed budget. To reduce the amount of time required for the participants, a set of six player options were provided for each of the five positions along with salary information for "purchasing" that player based on current DraftKings data. Thus, each session required participants to select five players, each from a different set of 6 available players (each with unique performance data and salary requirements), and do this while staying within a predefined budget. This task demands making tradeoffs between the value of spending money on certain positions (i.e., first base, pitcher, shortstop) versus others. For example, if the collaborating spends too much on the best hitter available, they may not have funds left to get a top pitcher. While no perfect strategy exists to optimize score, an effective strategy requires the group to make skilled and thoughtful tradeoffs.

During five consecutive weeks of the 2024 MLB season, forecasting sessions were conducted twice per week (excluding the "all-star break"). Groups of approximately 25 participants, each self-identified as a baseball fan, were engaged each week. All of the participants reported being familiar with fantasy sports challenges. Participants were tasked first with individually choosing players to craft a "personal roster" with their goal to maximize points scored. A bonus was awarded during each session to high performers to motivate individuals to maximize their performance. The full group of participants was then gathered within the online Thinkscape CSI platform and tasked with collaboratively choosing players through conversational deliberation, their collective goal to maximizing points scored. Again, a bonus for strong performance was awarded to motivate participants on this collaborative task. In all sessions, the test was modeled on the real-world DraftKings contest and used authentic data from the DraftKings fantasy baseball site.

To save time for the participants, four of the nine positions on the baseball roster were pre-selected before the session began. The participants were then tasked with selecting the remaining five while staying within a fixed budget that was provided to them. The order in which the positions (i.e., pitcher, first base, shortstop) were presented to the groups was randomized for deliberation and selection in each session.

For the CSI trials, the 25 person groups were automatically divided into 5 subgroups of 5 people (approx.). Each of the subgroups also included a Surrogate Agent as described above, enabling conversational content to pass freely among subgroups. In addition, half the CSI sessions employed a custom-created "Infobot" that was primed with expansive statistical data about current MLB players and MLB teams using the Assistants API from Open AI. The members of each subgroup could freely query the Infobot for information related to MLB, but nothing else. The queries were part of the standard chat deliberations such that all members of a subgroup could see the responses in real-time.

To ask the Infobot for information, participants simply mention InfoBot's name ("@infobot") in conversation. An example is shown in Figure 3: a real participant (Mango) asked the local Infobot to provide statistics for the set of Third Basemen currently under consideration. Infobot responds with statistics broken down by player for the last 15 games.

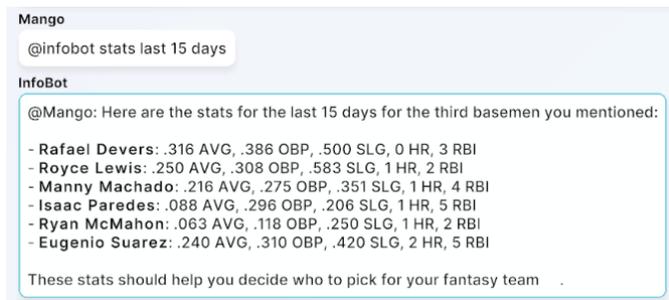

Fig 3. Infobot example in the Thinkscape application.

The deliberating group was provided a period of 5 minutes and 30 seconds to collaboratively select each player for each position. This resulted in sessions that lasted approximately 30-minutes to select all five positions. As the session progressed, the group was provided with an indication of their remaining budget. As a consequence, no group during the test fielded a roster that was overbudget as it was not allowed. The groups that used infobots had the same time per question (five minutes and 30 seconds) allotted for selecting each player.

To define a Wisdom of Crowd (WoC) roster of players based on the individual survey results, the following method was used: First, the most popular player choice for each position was used. Then, if the resulting roster would go over budget, the player chosen (of the five selected) with the lowest plurality of survey votes was replaced with the next most popular player choice for that position.

In addition, an exit survey was administered to all participants of CSI sessions and captured subjective feedback as to their perception of the CSI experience itself, and their perception of using the experimental Infobot capability.

IV. RESULTS

Data was collected for 10 sessions, each requiring a different set of players be chosen for a different set of MLB games. One of the two sessions each week used standard CSI, and the other session each week also was augmented with an Infobot. Every session was scored using official DraftKings scoring methods based on the official MLB results from the related games. The scoring for each session was only for the five positions that were selected by participants for that day's Daily Fantasy contest.

The collaborative rosters that were generated using the CSI platform scored 62.4 points per session. As indicated in Table 1, this method outperformed the median individual's score (47.3 points) on personal rosters (generated by survey). A paired t-test by session was performed and shows that this difference is statistically significant (p=0.004). The group that deliberated using CSI also significantly outperformed the scores generated using the WoC method (i.e., by taking most popular selection across the set of individual surveys), which averaged 43.7 points across the ten weeks (p=0.008).

| Aggregation Method | Average Points | Average Percentile Performance |
|---|---|---|
| Thinkscape (CSI) | 62.4 | 73.2% |
| Survey (WoC) | 43.7 | 38.6% |
| Survey (Median Individual) | 47.3 | 50.0% |

Table 1. Performance Comparison in MLB Fantasy Contest

An alternate way to compare performance is to look at the percentage of individual survey participants that each method outperformed. On average, groups using the CSI platform exceeded the score of **73%** of individually generated rosters. This was significantly more accurate than the WoC method which outperformed **39%** of individually generated rosters (p-value = 0.011) and greatly exceeded the Median individual score (p-value = 0.016) using a Wilcoxon Rank Test for the points scored by lineup choices. Additionally, we are 99% confident that the true percentile performance of CSI-based sessions is between 69% and 77%. In addition, a bootstrap test was performed by resampling 10,000 times over the observed participant scores. This test indicated that the CSI method is likely to outperform the majority of individuals 95% of the time.

Moreover, we compared the sessions that used the Infobot to those that did not. As shown in Table 2 below, there was a small improvement in performance for the Infobot sessions, but it was not statistically significant. Table 2 also compares the average number of characters per minute in each subgroup across all sessions. As shown, the conversations were measurably more efficient (p=0.012) when using Infobots (183 characters per min) versus without Infobots (197 characters per min).

| Infobot Condition | Average Percentile Performance | Characters Per Minute (Subgroup) | Conversational Variance per participant |
|---|---|---|---|
| Thinkscape (CSI) without Infobot | 72% | 197 | 15.0 |
| Thinkscape (CSI) with Infobot | 74% | 183 | 12.2 |

Table 2. Comparing Infobot vs No Infobot Sessions

Examining the variance in characters per minute per participant for each question, we find that there is significantly less variance (p=0.039) between participants in each subgroup when the Infobot was used (12.2 characters per minute) as compared to when the Infobot was not used (15.0 characters per minute). This may be because the Infobot is an "equalizer" among participants, giving those who have less factual knowledge about MLB an easy to way consider the player rigorously and make confident comments that they may not have made if they felt insecure in their expertise compared to other members of their subgroup.

To appreciate how CSI deliberations foster superior performance versus traditional WoC aggregation, the following example describes one deliberation during a real session. In this example, the group is down to their final position selection of the day and is collaboratively considering who to pick to play Second Base in their roster. The six choices available to them are shown in Table 3 below along with the cost of purchasing each player. The group's remaining budget is $4,900. From a reputation perspective, Marcus Semien is considered the best player in the set, but he is also the most expensive. That said, the group had sufficient budget left to cover this cost, so he would be the common pick for most participants. In fact, the survey results showed a strong majority picking Semien.

| 2B | Marcus Semien | $4,900 |
|---|---|---|
| 2B | Xander Bogaerts | $4,800 |
| 2B | David Hamilton | $4,500 |
| 2B | Brendan Rogers | $4,400 |
| 2B | Gavin Lux | $3,500 |
| 2B | Jackson Holliday | $3,400 |

Table 3. Set of players that group was tasked with choosing from.

Despite the conventional wisdom supporting Semien, the CSI conversation yielded a different result. Figure 4 shows the time-varying "sentiment" towards each of the six MLB players across all give subgroups (i.e. all 25 participants). As shown, the deliberation began similarly to the survey results, with sentiment quickly surging for Marcus Semien as the best selection. This continued for most of the deliberation until an insight was raised within one of the subgroups – the fact that Brendan Rogers was on an unusual hot streak over the last few games. As this insight spread among subgroups, the sentiment in favor of Rogers rapidly gained traction and the deliberation ended with the group picking Rogers for the final slot in their roster, defying the conventional wisdom.

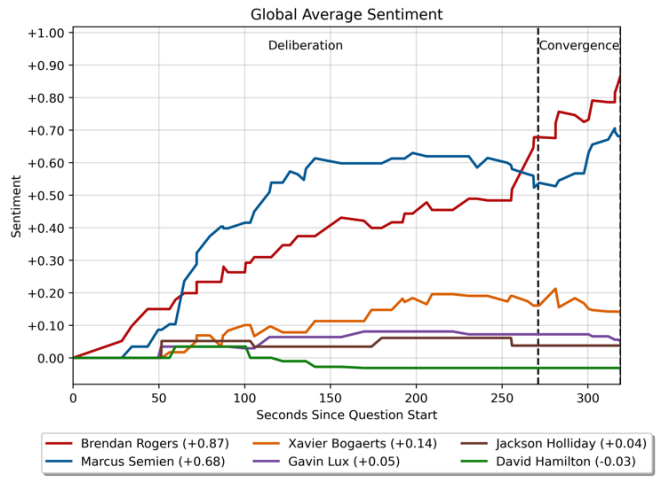

Fig 4. Time-varying Conversational Sentiment across subgroups of a 25-person deliberation using CSI

This example shows how CSI enables thoughtful dialog on a local scale (in small subgroups) while also allowing unique insights to spread across the full population and influence the global decision. In this case, the actual results of this non-obvious pick have interesting anecdotal value. That's because of all six players the group could have chosen for that Fantasy Baseball session, the one that scored the highest when the games were played was in fact Brendan Rogers. On the other hand, the obvious pick by aggregating conventional wisdom, Marcus Semien, underperformed expectations. Ultimately it is because of this type of dynamic in a variety of questions across the 5-week study that enabled the CSI groups ( with and without Infobots) to outperform most individuals and outperform the WoC aggregations.

5.1 Subjective Feedback Surveys

In addition to assessing the performance of CSI trials in the MLB Fantasy contest, subjective feedback was collected from all participants about their experience selecting players via CSI deliberations. These surveys were aggregated across all 10 sessions, each including approximately 25 respondents. As shown in Figure 5, Tables 4 and Table 5, over 90% of respondents agreed with each of the positive statements about CSI tested, and over 80% of respondents disagreed with the negative statements about CSI tested. Taken together, these results suggest that participants in CSI deliberations felt strongly that their perspectives were heard and considered by others, and that group benefited from the CSI deliberations and converged thoughtfully on strong solutions without rushing to conclusions. Thus, these CSI groups did not suffer from low psychological safety, or the sense that offering opinions will be met with negative judgements, commonly found in deliberating groups.

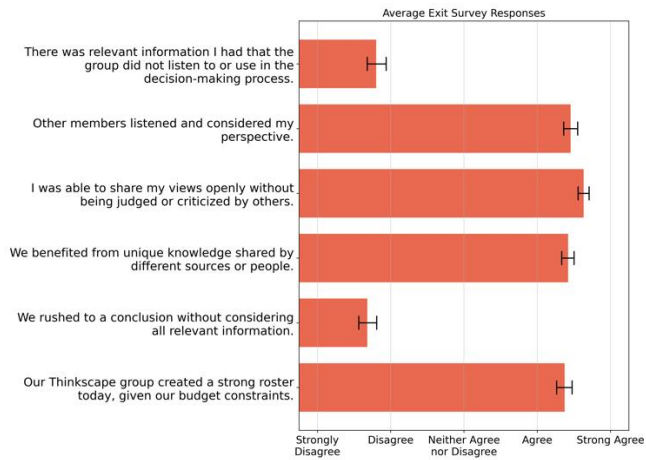

Fig 5. Subjective Survey Results across all 10 sessions

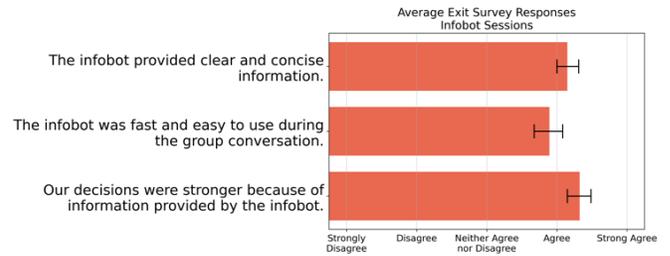

Fig 6. Additional Subjective Survey Results from Infobot sessions

| Statement | Percentage of Agreement |
|---|---|
| Other members listened and considered my perspective. | 93.9% |
| I was able to share my views openly without being judged or criticized by others. | 97.3% |
| We benefited from unique knowledge shared by different sources or people. | 97.3% |
| Our Thinkscape group created a strong roster today, given our budget constraints. | 93.2% |

Table 4. Subjective Feedback Results on Positive Statements

| Statement | Percentage of Disagreement |
|---|---|
| There was relevant information I had that the group did not listen to or use in the decision-making process. | 81.6% |
| We rushed to a conclusion without considering all relevant information. | 87.8% |

Table 5. Subjective Feedback Results on Negative Statements

In addition, the participants in sessions that used Infobots were asked three additional subjective questions about the use of Infobots. As shown in Figure 6 and Table 6, over 70% of respondents agreed or strongly agreed with all three of the statements provided about Infobots use. These results suggest that participants found the Infobots to be easy to use and to provide clear and concise information. In addition, over 86% of participants either agreed or strongly agreed with the assertion that their collective decisions were "were stronger because of information provided by the Infobot."

| Statement | Percentage of Agreement |
|---|---|
| The Infobot provided clear and concise information. | 82.4% |
| The Infobot was fast and easy to use during the group conversation. | 74.3% |
| Our decisions were stronger because of information provided by the Infobot. | 86.5% |

Table 6. Subjective Feedback Results on Infobot Statements

In addition to positive feelings about using the Infobot, the actual usage statistics were tracked and aggregated across the five sessions that used the intervention. These results show that participants queried the Infobots regularly for factual information during all 25 questions in which Infobot was present. Although the groups were not required to use the Infobot to help in deliberation, an average of 4.1 Infobot queries were made per subgroup per player being selected. The consistency of use of the Infobot is also of note, with all questions averaging between 2.8 and 5.5 queries per subgroup. The fact that all subgroups consistently made strong use of the infobot without being required to do so, suggests strong value from this feature during group deliberations.

## V. CONCLUSIONS

To test the performance benefits and the perceived value of real-time conversational deliberation using a CSI platform, we conducted a collaborative forecasting study based on a well-known daily Fantasy Baseball contest. This task is open-ended, complex, and requires that participants stay within a fixed budget across multiple player sections. To perform well, collectives needed to weigh the benefits of picking the highest performing players for certain positions against the benefits of saving budget for other positions on the roster.

The results showed that when using CSI, 25-person groups outperformed 72% of individually surveyed participants and showed significant intelligence amplification versus the mean scoring individual respondent (p=0.016). The CSI groups also outperformed the most popular picks across sets of 25 of surveys for each daily contest (p<0.001). This suggests that real-time deliberative conversation is a superior method for harnessing the collective forecasting power of a 25 person group than aggregating survey responses.

In addition, this study tested an additional intervention in half of the CSI sessions – the deployment of an informational AI assistant called an Infobot. Each conversational subgroup of 5 participants had access to their own Infobot and used it frequently (on average, 4.1 queries per player being selected). There was not a significant difference in scoring between the CSI sessions with and without Infobots, but the feedback from participants was very positive. 85% of participants agreed with the statement "Our decisions were stronger because of information provided by the Infobot," and only 4% disagreed. In addition, deliberations in subgroups using Infobots showed significantly less variance ($p=0.039$) in conversational content. This suggests that Infobots promoted more balanced discussions in which fewer members dominated the dialog. This may be because the infobot enabled participants to confidently express opinions on players that they might not have known a lot about as a consequence of asking the Infobot for factual information. Finally, participants reported that they could speak freely without fear of negative interpersonal repercussions, which suggests the CSI platform and Infobot AI assistant facilitated open discussions without falling victim to common conversational issues found in large groups.

ACHNOWLEDGEMENTS

This research and development project was funded in part by the DOD/USAF under contract FA864924P0820.